# A Smart Adaptively Reconfigurable DC Battery for Higher Efficiency of Electric Vehicle Drive Trains

Zhongxi Li, Aobo Yang, Gerry Chen, Nima Tashakor, Zhiyong Zeng,
Angel V. Peterchev, Stefan M. Goetz

*Abstract*—This paper proposes a drive train topology with a dynamically reconfigurable dc battery, which breaks hard-wired batteries into smaller subunits. It can rapidly control the output voltage and even contribute to voltage shaping of the inverter. Based upon the rapid development of low-voltage transistors and modular circuit topologies in the recent years, the proposed technology uses recent 48 V power electronics to achieve higher-voltage output and reduce losses in electric vehicle (EV) drive trains. The fast switching capability and low loss of low-voltage field effect transistors (FET) allow sharing the modulation with the main drive inverter. As such, the slower insulated-gate bipolar transistors (IGBT) of the inverter can operate at ideal duty cycle and aggressively reduced switching, while the adaptive dc battery provides an adjustable voltage and all common-mode contributions at the dc link with lower loss. Up to $^2/_3$ of the switching of the main inverter is avoided. At high drive speeds and thus large modulation indices, the proposed converter halves the total loss compared to using the inverter alone; at lower speeds and thus smaller modulation indices, the advantage is even more prominent because of the dynamically lowered dc-link voltage. Furthermore, it can substantially reduce the distortion, particularly at lower modulation indices, e.g., down to $^1/_2$ compared to conventional space-vector modulation and even $^1/_3$ for discontinuous pulse-width modulation with hard-wired battery. Other benefits include alleviated insulation stress for motor windings, active battery balancing, and eliminating the vulnerability of large hard-wired battery packs to weak cells. We demonstrate the proposed motor drive on a 3-kW setup with eight battery modules.

*Index Terms*—electric vehicle, battery, pulsating dc-link, shared modulation, converter losses.

## I. Introduction

CONCERNS about the environmental impact and future shortage of fossil fuels is driving the growths of electric vehicle (EV) fleets. Their drive trains almost exclusively use hard-wired battery packs and two-level three-phase inverters due to their simplicity, compactness, and ruggedness [1]–[5].

However, the simplicity entails several critical problems. First, the dc-link voltage is uncontrolled and rather high at all times regardless of which voltage the motor requires in real time, imposing maximum switching and insulation [2], [6], [7]. Furthermore, hard-wired batteries in EVs are high-voltage sources, even when the battery is empty. Finally, modulation techniques to reduce the switching loss, such as flat-top or discontinuous modulation, are known but reduce the quality and harmonic content of the output with substantial impact on losses and degradation of the motor [8], [9].

The constant presence of high voltage is not only already complicating the assembly in factories—which necessities skilled workers with high-voltage qualifications—but is regularly confusing first responders at car crash sites leading to possibly delayed rescue [10]–[14]. The safety concerns are further aggravated by the recent trend toward higher battery voltages [5]. It would be highly advantageous being able to dynamically and reversible break up the massive high-voltage batteries (often with 100 to 250 cells in series) in such situations.

Another design challenge entailed by the hard-wired battery concept is the large variation of the output voltage due to load and state of charge (SoC). In consequence, all loads have to operate at the entire voltage range, ideally such that the driver of a vehicle does not recognize an SoC influence on performance. Consequently, low SoCs should not affect the speed, and high SoCs should not exceed any ratings. The selection of the transformer ratio in isolated dc/dc converters, e.g., for auxiliary supply particularly suffers from large input voltage spreads. The consequence is a large compromise with respect to size, cost, and efficiency [15], [16]. The battery voltage is practically far from the ideal level at any point in time. Due to the wide range and increasing trend of voltages of hard-wired batteries in EVs, previously installed dc chargers on the street are gradually losing compatibility [17]. Manufacturers therefore offer high-power built-in dc/dc stages to adapt to the charging voltage but obviously increases the cost and weight [5], [18], [19].

Furthermore, the vehicle industry is struggling with another effect associated with hard-wired batteries, which is aggravated with latest 800 V vehicle architectures [5], [20], [21]: production tolerances let newly manufactured battery cells spread with respect to available capacitance, self-discharge, heating, maximum power capability, and ageing. One might assume that such spread averages out in large packs with several hundred cells, but the opposite is true: once the first cell is empty or critically hot, the entire pack has reached its limits, and discharging has to stop—no matter how well or full other cells still are. When the first cell is broken, the entire pack is out of service. Assuming a certain Gaussian spread for each cell, the likelihood for a pack to contain an outlier far below the average performance statistically grows with size. Thus, the larger the battery pack becomes, the poorer is the usable performance. Costly cell balancing, which is mandatory to not overcharge cells, does not solve the limitation by the weakest cell with respect to capacitance, power, or ageing [16].

Many of the above problems can be solved by adjusting the dc-link voltage, e.g., regulating the dc-link voltage with respect to the speed. A dc/dc converter can, for instance, mitigate the large variation of battery voltage with SoC and the extreme mismatch between available and ideal voltage with respect to efficiency. Some vehicles, branded commercially as hybrid synergy drive by Toyota, implement a dc/dc converter between the hard-wired battery and the inverter to compensate the effect.

In such vehicles, the voltage of the dc link is controlled typically via boost, rarer a buck or buck-boost dc/dc stage between the battery pack and the three-phase inverter [6], [7], [16], [22]. At low speeds, the lowered dc-link voltage alleviates the switching loss in the inverter and notably improves the total efficiency. At higher operating speeds, however, the dc/dc stage is less beneficial to the frontend inverter while producing noticeable extra losses itself [2]. The dc/dc stage also requires large passive components, which impede efforts to reduce weight [23]–[25]. Furthermore, the boost topology, though mitigating the battery current ripple, in combination with the constant-power control of the drive inverter entails known stability issues and tends to further enlarge filters for decoupling [26], [27]. More importantly, the other problems of hard-wired battery packs, including safety and the limiting impact of the weakest cell, cannot be solved by the dc/dc stage.

Both research and the industry are developing and driving a number of alternative integrated battery–power-electronics combinations in lab setups and concept vehicles that would solve the issues by modularizing or even reconfiguring the previously hard-wired battery more or less during operation [12], [28]–[38]. Various reconfigurable battery technologies are in product development at several companies, demonstrating their economic potential [30], [39]–[47].

Recently presented drive-train designs with a reconfigurable dc battery adjust the steady-state dc-link voltage for increased efficiency and stability, while mitigating the limitation of conventional hard-wired battery packs on the weakest cell [27]. With latest circuit topologies, reconfigurable batteries can concurrently generate multiple output voltages and supply various auxiliaries below the drive-train voltage [48].

Subunits with 12–15 cells are often selected as the reconfiguration level to reach a good trade-off between granularity and signaling/driving effort. Such subunits also stay below the safe extra-low voltage (SELV) of 60 V. The reconfiguration of small sub-60 V blocks instead of large hard-wired battery packs can in principle solve all above-mentioned issues; residing on the previously present module management and balancing electronics within each module and using the existing battery cooling system, it requires low extra cost and space, which is overcompensated by the advantages. The limitation by weak cells in growing battery packs is naturally circumvented as outliers only limit their subunit. The modularization in combination with dynamic reconfiguration allows deep battery analysis using the already present transistors [49]. Likewise, passive balancing concentrates on the sub-unit level, while circuit reconfiguration of the subunits allows trading off all discrepancies in power, capacitance, heating, and ageing [50].

However, the concepts presented so far exclusively operate the battery and dc-link voltages quasi-stationary with only very slow adjustments [27]. The high-speed capability of the FETs in the reconfigurable battery is hardly used and the transistors only rarely switched despite an unprecedented development of such transistor technology in recent years due to their need in 48 V vehicle systems, consumer electronics, and white goods [51], [52]. This paper illustrates how a system-level perspective and smarter integration of such adaptive dc batteries would allow distributing the task of motor drive over inverter and battery for the sake of higher overall efficiency and quality. We recently presented the idea with a preliminary status at IECON, but with only very limited theoretical analysis, little consideration of filtering as well as practical design aspects, and steady-state operation data [53]. This paper wishes to fill this void and is organized as follows: Section II details the approach in response to above need and relates it to the prior art. Section III introduces the topology and the modulation principle for the drive inverter. Section IV discusses the modular multilevel dc battery. Section V quantitatively compares the proposed solution with existing technology. Section VI presents experimental results.

## II. Concept of the Adaptive DC Battery

### A. Adaptive DC Battery Through Circuit Reconfiguration

This paper presents a cascaded module or modular multilevel implementation of an adaptive dc battery (named *backend* in the following, see Fig. 1) that is actively involved in the motor drive control. The battery sub-units interface each other through cascaded half-bridges (CHB) or double half-bridges (CHB²) and together feed a two-level three-phase inverter (*frontend* in the following, see Fig. 1). CHB or CHB² modules break the large total battery voltage into smaller units so that both the subunits of the battery and the transistors only need to manage a small portion and can be low voltage. The multilevel adaptive dc battery allows (1) ***flexible active battery management***, including charge and power balancing, without external circuits, (2) ***quasi-stationary dc-link voltage adjustment*** that follows the motor speed, and (3) ***shared modulation*** between the adaptive dc battery and the inverter without large passive components

Specifically, the adaptive dc battery produces a six-pulse rectified waveform at the dc-link to dynamically shape the modulation region for the frontend inverter and consequently keep two of the three inverter phases in a fixed switching position. In equivalent terms, no zero-voltage vectors are required. The frontend inverter therefore spares $^2/_3$ of switching compared to conventional space-vector pulse-width modulation (SVPWM), or $^1/_2$ compared to discontinuous pulse-width modulation (DPWM) and still without the distortion issues of DPWM [25], [54], [55]. The spared switching duty is relegated to the adaptive dc battery backend, but the latter incurs much less switching loss due to the fractionized switching voltage and the use of field-effect transistors (FETs) and can generate higher output quality due to a higher number of output levels.

As will be shown in the later sections, the proposed configuration notably reduces the overall loss compared to a conventional setup with fixed hard-wired battery dc-link. The advantage is most prominent at lower modulation indices (emulating low speeds and smaller back emf) or at lower power factors (prominently in field-weakening operation at high speeds). The saved loss mostly owes to the reduced switching loss of the insulated gate bipolar transistors (IGBTs) in the frontend inverter, which cannot easily benefit from more silicon in parallel; the reduced switching loss can be traded for better conduction loss in the IGBT selection [56]–[58]. Furthermore, the output quality of the proposed converter remains good across all modulation ranges whereas SVPWM and DPWM with fixed dc-link voltage inevitably suffer higher distortion at lower modulation indices. The proposed topology is demonstrated on a 3 kW setup with eight modules at the battery backend and a two-level three-phase IGBT module at the frontend inverter.

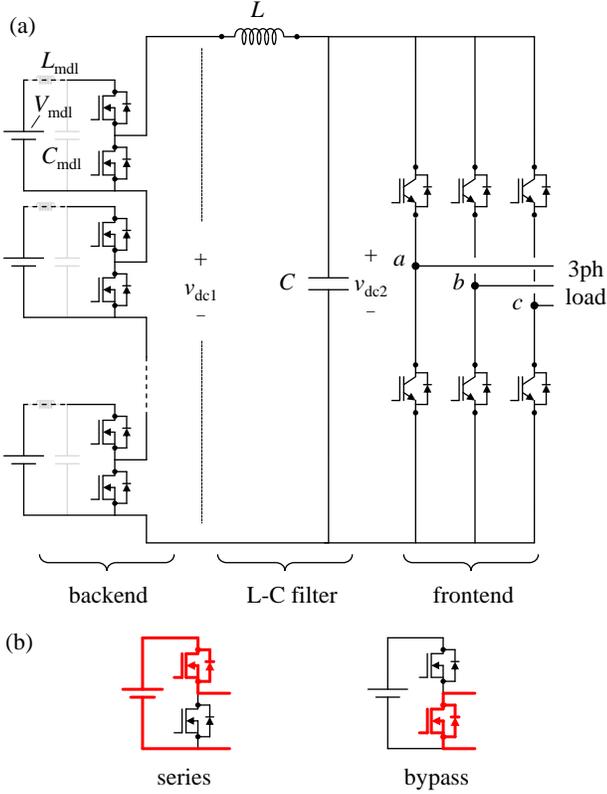

Fig. 1. (a) Proposed three-phase pulsating dc-link motor drive. A small L-C filter is implemented between dc-links $v_{dc1}$ and $v_{dc2}$. Each module can incorporate an optional battery filter ($L_{mdl}$ and $C_{mdl}$). (b) Switching states of the battery backend, where the battery filters are ignored.

*B. Relation to AC Batteries*

Similar to the proposed solution, so-called ac batteries are intensively studied as they similarly provide excellent output quality, very low voltage transients d$v$/d$t$, and flexible management of the energy storage elements [59]–[68]. Most of these ac battery approaches require a relatively high number of battery cells in series to generate all phases of the high voltage required for three-phase motors. However, smaller and more cells as required for most three-phase ac batteries go against the trend of the vehicle industry to larger battery cells for higher system-level energy density due to lower overhead from packaging material, as is evidenced by recently released and announced EV models [69], [70]. More importantly, ac batteries are typically based on cascaded full bridges (CFB), which lead to a low battery utilization and further load the battery subunits with large reactive power, such as second-order ripple current with an amplitude as high as the active component [71]–[78]. Substantial effort in topologies and control can mitigate those problems [66], [67], [71], [79]. In comparison, the proposed system of adaptive dc battery and inverter does not suffer from second-order ripple currents thanks to trigonometric cancellation while retaining the other advantages of reconfigurable battery circuits, such as flexible battery management and in contrast to hard-wired batteries achieve the optimum performance with respect to capacitance, power, and ageing. Since the proposed topology only needs one unipolar module string with half-bridge electronics, it uses 70% fewer[1] transistors compared to bipolar three-phase ac batteries with CFB at the same output voltage rating.

*C. Relation to Previous DC-Link Modulation Methods*

The concept of quasi-stationary dc-link voltage adjustment was studied with monolithic dc/dc stages, reflecting the topology of Toyota's hybrid synergy drive and other in-production vehicles [54]. A modulation of the dc voltage can substantially reduce the switching loss in the subsequent drive inverter [7], [25], [54], [55], [80], [81]. However, the monolithic dc/dc stage still switches the full system voltage and current, requiring high-voltage, high-current components, suffers from a trade-off between control instabilities and dynamics, and does only solve some of the issues discussed in Section I.

The concept of shared modulation was also studied previously. Raju and colleagues as well as Klumpner and Zargar, for instance, insert a full-bridge circuit between the backend dc link and the frontend inverter to shape the dc-link waveform—the use of the full-bridge relegates the high-voltage switching to low-voltage parts [54], [55], [80]. However, when used for EV drive trains with batteries, those topologies only offer a narrow dc-link voltage range, losing the ability for adjusting the overall voltage to the needs of the drive, although that was one of the initial motivations of dc-link modulation. Furthermore, despite the extra semiconductors, the circuit does not offer flexible active battery management since there is no additional controllability over battery subunits compared to a hard-wired battery pack.

Alternatively proposed topologies use buck or boost stages [7], [25]. However, the transistors in the buck or boost stages must match the system's maximum voltage and current ratings. Hard-switched dc/dc stages are typically not considerably more efficient than switching in the inverter, which topologically is a combination of buck legs [82]. Consequently, the loss reduction is at best marginal. Larger magnetic components, which have to deal with the full load current, are also mandatory in a buck or boost stage, imposing a trade-off between weight and efficiency.

### III. TOPOLOGY AND OPERATING PRINCIPLE

Fig. 1 shows the proposed three-phase motor drive. It contains a modular reconfigurable battery as the backend, a three-phase two-level inverter as the frontend, and a small L-C filter in between. The adaptive dc battery in the backend generates the intermediate product vdc1, which is further filtered by the L-C filter to produce $v_{dc2}$ to form the rectified common-mode of the motor phases. Voltage $v_{dc2}$ is shared with the frontend drive inverter. The L-C filter is tuned to decouple the switching transients between the frontend inverter and the adaptive backend battery, and therefore can be small.

---

[1] A factor of ½ due to the unipolar modules, together with an additional factor of √3/3 due to fewer arms, specifically only one instead of three for three-phase ac output.

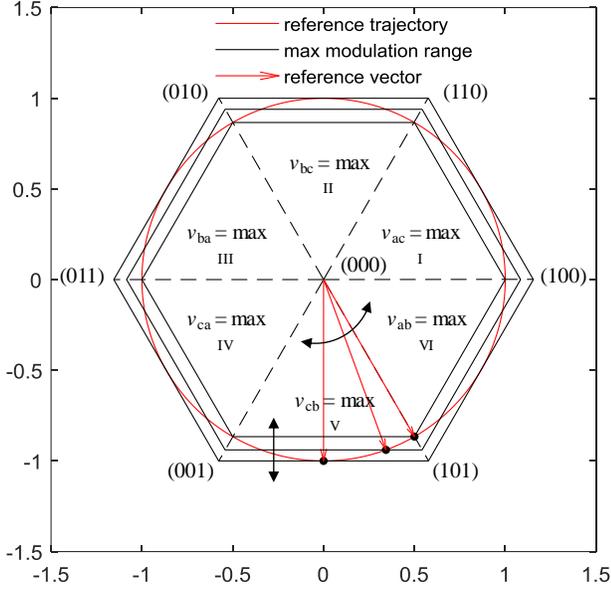

Fig. 2. Equivalent space-vector diagram. Black arrows indicate progression in time.

TABLE I
Activated sectors and modulation references

| max line voltage | $v_{ac}$ | $v_{bc}$ | $v_{ba}$ | $v_{ca}$ | $v_{cb}$ | $v_{ab}$ |
|---|---|---|---|---|---|---|
| sector # | I | II | III | IV | V | VI |
| $m_a$ | 1 | $v_{ac}/v_{dc2}$ | 0 | 0 | $v_{ab}/v_{dc2}$ | 1 |
| $m_b$ | $v_{bc}/v_{dc2}$ | 1 | 1 | $v_{ba}/v_{dc2}$ | 0 | 0 |
| $m_c$ | 0 | 0 | $v_{ca}/v_{dc2}$ | 1 | 1 | $v_{cb}/v_{dc2}$ |

For introducing the operating principle, let us first ignore the battery backend and assume that $v_{dc2}$ can be freely controlled. From the inverter's point of view, the optimal waveform for $v_{dc2}$ is

$$v_{dc2} = \max\{v_a, v_b, v_c\} - \min\{v_a, v_b, v_c\} \\ = \max\{v_{ab}, v_{bc}, v_{ca}, -v_{ab}, -v_{bc}, -v_{ca}\} \quad (1)$$

where $v_a$, $v_b$, $v_c$ are reference phase voltages. Equation (1) is optimal in the sense that two inverter phases remain unswitched at any moment, regardless of the references. The low switching effort is a consequence of $v_{dc2}$ tracking the maximum output voltage envelope per (1).

The reduction in the switching is clear when viewed in the space-vector diagram in Fig. 2. The circular reference trajectory is inferred from upper-level controls. The hexagon(s) reflect the momentary modulation range (defined by $v_{dc2}$) and therefore have variable size. Equation (1) sets $v_{dc2}$ in such a way that the hexagon always coincides with the momentary reference vector (Fig. 2) and widely eliminates the amplitude modulation for shaping the ac output, effectively leaving mostly commutation to the frontend inverter. Without the fine amplitude modulation, the reference can be conveniently synthesized by the two adjacent active vectors without zero vectors, and only one phase leg switches at high frequency. Compared to the conventional SVPWM scheme, the elimination of the zero vector spares $2/3$ of switching. Precise evaluation of the overall loss depends on the load angle, output current, as well as the loss of the backend adaptive dc battery, which are detailed in Section V. In short, the overall loss turns out lower because regulating $v_{dc2}$ takes much less effort using the backend FETs than inserting zero vectors with the frontend IGBTs. Furthermore, for power factors larger than 0.5, the frontend inverter never switches during current apexes.

Control of the frontend inverter requires *1)* identifying the active sector and *2)* modulation between the two adjacent active vectors. The sector is identified by the momentarily largest line voltage, as is labeled in Fig. 2. For instance, $v_{ab}$ = max activates Sector VI, whereas $v_{ba}$ = max activates Sector III. The sector determines which half-bridge to modulate. The gate signal is obtained by comparing a reference with a unipolar triangle carrier. TABLE I lists the modulation references according to the activated sectors, which are further unified below [54]:

$$m_x = \frac{v_x - \min\{v_a, v_b, v_c\}}{\max\{v_a, v_b, v_c\} - \min\{v_a, v_b, v_c\}}, x = a,b,c. \quad (2)$$

## IV. ADAPTIVE BATTERY BACKEND

In contrast to previous approaches with monolithic dc/dc stages and other separate units [7], [25], [54], [55], [80], [81], we use a modular adaptively reconfigurable multilevel dc battery as the backend connected to the frontend inverter through a dc link (Fig. 1). The dc battery is implemented with a cascaded structure where each module needs to be rated only for a fraction of the total voltage and incorporates a battery unit. As such, low-voltage high-current FETs can be used. The adaptive dc battery's physical modularity is perfectly suited for this pulsating-dc-link EV drive mainly because of *1)* flexible control over individual batteries and thus extended mileage [83]–[85] and *2)* excellent output quality while causing negligible switching loss [86]–[90]. Importantly, we will show that the high-fidelity output allows a much smaller L-C filter and thus a rapid response.

Fig. 1(b) lists the switching states. The adaptive dc battery voltage $v_{dc1}$ is determined by the number of series states,

$$v_{dc1} = n_{series} V_{mdl}, \quad (3)$$

where $V_{mdl}$ is the voltage of the battery subunits, $n_{series}$ is the number of modules in the series state.

We use a phase-shifted carrier (PSC) scheme to control the output, modulate the adaptive dc battery, and balance the modules. Specifically, the PSC scheme assigns a carrier $C_k$ ($0 \leq C_k \leq 1$) to the $k$-th battery module. The switching state is determined by

$$\text{state}(k) = \begin{cases} \text{series}, & \text{if } m_{dc} \geq C_k, \\ \text{bypass}, & \text{if } m_{dc} < C_k, \end{cases} \\ m_{dc} = \frac{\max\{v_a, v_b, v_c\} - \min\{v_a, v_b, v_c\}}{N_{mdl} V_{mdl}}, \quad (4)$$

where $N_{mdl}$ denotes the number of battery modules.

We adapted previous work allowing individual module balancing in cascaded double H bridges using phase-shifted carriers for ac and dc output [91]–[93]. For reconfigurable batteries based on cascaded modules in general, battery balancing can be achieved without additional hardware or losses by superimposing small dc values $\Delta m_{x,k}$ to the modulation indices [94], [95]. As long as $\Sigma \Delta m_{x,k} = 0$, there is no influence on the external load. The differential load for the $k$-th module of phase $x$ is

$$\Delta p_{x,k} = \Delta m_{x,k} V_{mdl} i_x \quad (5)$$

where $i_x$ denotes the current of phase $x$.

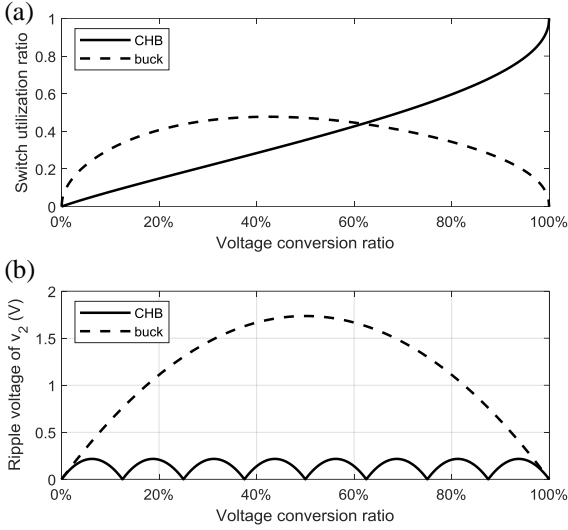

Fig. 4. (a) Switch utilization ratios (SUR). (b) Estimated ripple voltage of CHB and buck-dc/dc backend converters. $L = 30\ \mu H$, $C = 60\ \mu F$, $N_{mdl} = 8$, $f_s(\text{buck}) = N_{mdl} f_s(\text{CHB}) = 40$ kHz, and $V_{mdl} = 40$ V. Any influence of the frontend inverter is neglected.

## V. ANALYSIS AND COMPARISON

This section compares the proposed motor drive with alternatives. To show the advantage of the variational dc-link voltage, we compare the proposed solution with a conventional single-stage two-level three-phase inverter emphasizing on converter losses and total harmonic distortions (THDs). To demonstrate the benefits of the variable dc battery backend concept, we compare it with a buck converter with a focus on the switch utilization ratio (SUR) and quality of the dc-link waveform.

### A. Cascaded Variable DC Battery Backend vs. Buck Converter

The reconfigurable dc battery backend essentially operates as a multilevel step-down dc/dc converter. We compare its switch utilization ratio (SUR) with that of a buck converter according to [96]

$$\text{SUR} = P_{\text{load}}/S, \quad S = \Sigma V_j I_j, \quad (6)$$

where $S$ is the *total active switch stress*, defined as the sum of the products of the peak voltage $V_j$ and rms current $I_j$ seen by each switch. The SURs are shown in Fig. 4 with various voltage conversion ratios. The SUR of the variable dc battery is advantageous in the high modulation region, which corresponds to approximately the range from 25 mph/40 kph up in many vehicles with single-speed gearbox.

### B. Voltage Ripple Caused by the Backend Converter

The SURs reflect the conduction loss with the same semiconductor budget. Further comparisons should evaluate the quality of $v_{dc2}$ under the same effective switching frequencies.

We assume $N_{mdl} V_{mdl}$ at the input of the buck converter, matching the maximum voltage of the adaptive dc battery. With the same L-C filter, the two topologies produce the following voltage ripple at $v_{dc2}$:

TABLE II
Simulation settings

| | | |
|---|---|---|
| **Shared parameters** | | |
| Output power | $P$ | 100 kW ($m = 1$) |
| Output current | | 200 A pk |
| Output frequency | $f$ | 50 Hz |
| Frontend inverter | | IGBT: 300 A/1200 V (FF300R12ME4†, Infineon) |
| IGBT gate driver | | +15 V /−5 V ideal, $r_g = 1.8\ \Omega$ |
| Frontend inverter switching frequency | $f_{inv}$ | 10 kHz (i.e., 3.3 kHz per switch) |
| **Case 1: Proposed modular multilevel dc-link motor drive** | | |
| Backend modules | | 40 V × 16 |
| DC-link voltage | $v_{dc2}$ | 0–640 V adjustable |
| Module switch | | FET: 0.75 mΩ, 300 A/60 V (IPT007N06N‡, Infineon) |
| Module switching frequency | $f_{mdl}$ | 5 kHz (i.e., 90 kHz at $v_{dc1}$) |
| FET gate driver | | +15 V /−5 V ideal, $r_g = 1.8\ \Omega$ |
| Frontend reference | $m_x$ | Eq. (2) |
| Backend reference | $m_{dc}$ | Eq. (4) |
| **Cases 2 and 3: Two-level three-phase inverter** | | |
| DC-link voltage | $v_{dc}$ | 640 V fixed |

† Provided in PSpice default library.
‡ SPICE model downloadable from www.infineon.com.

(a) Modulation index = 95%

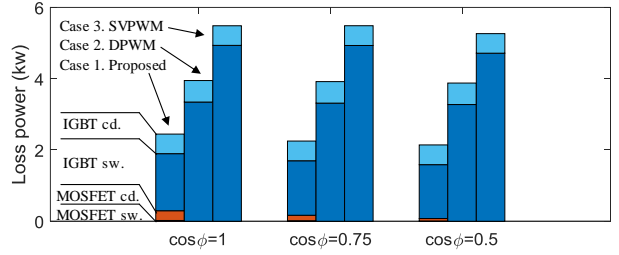

(b) Modulation index = 75%

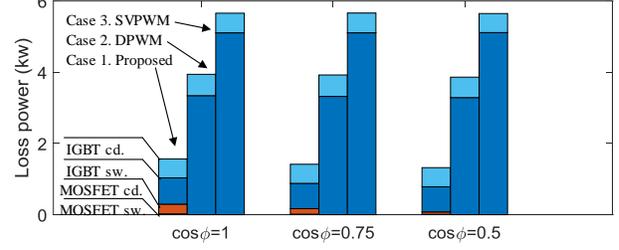

(c) Modulation index = 50%

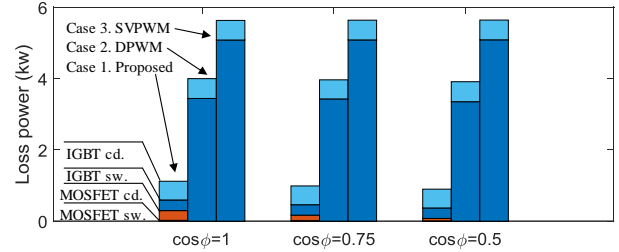

Fig. 3. Loss breakdown of three different cases under (a) 95% modulation index, (b) 75% modulation index, and (c) 50% modulation index. The switching loss of the FETs is barely visible.

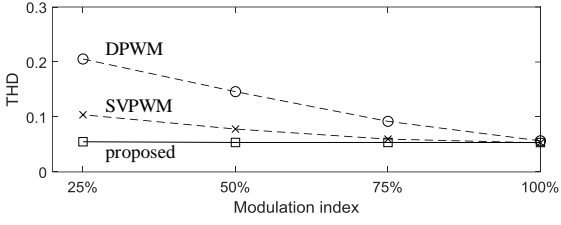

Fig. 5. THDs under series R-L passive load (1.75 Ω/200 μH). The modulation index refers to the utilization of the maximum dc-link voltage. See TABLE II for other properties. The modulation carrier frequencies are identical across three cases.

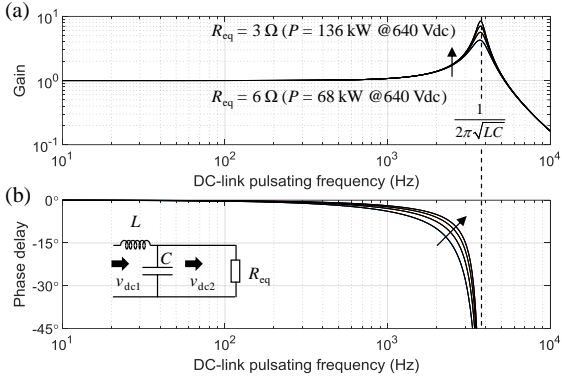

Fig. 6. Gain and phase shift across the L-C filter at the dc link. The dc-link pulsating frequency is six times of the electrical output frequency.

$$\Delta v_{dc2}(\text{CHB}) = \frac{V_{mdl}\left(\lceil m_{dc}N_{mdl}\rceil - m_{dc}N_{mdl}\right)\left(m_{dc}N_{mdl} - \lfloor m_{dc}N_{mdl}\rfloor\right)}{16LCf_s^2 N_{mdl}^2},$$

$$\Delta v_{dc2}(\text{buck}) = \frac{N_{mdl}V_{mdl}(1-m_{dc})m_{dc}}{16LCf_s^2}.$$

(7)

The numerical results are shown in Fig. 4(b) with the parameters of TABLE II. The effective switching frequencies of both topologies are matched for fairness. The adaptive dc battery produces substantially smaller voltage ripple because of the fractioned switching voltage. As such, the adaptive dc battery can use a much smaller L-C filter and/or switching frequency than that of a buck backend.

### C. Loss Comparison

We analyzed the performance with SPICE-level component models from the manufacturers and included switching transients in PSpice with the parameters of TABLE II. We compare the losses across three cases: (1) the proposed modulated dc link, (2) conventional inverter with SVPWM, and (3) conventional inverter with discontinuous PWM (DPWM) [25], [54], [55]. In SVPWM, the transistors are constantly switching, whereas DPWM injects a special common-mode modulation reference, e.g.,

$$m_{com}^{(\text{DPWM})} = \begin{cases} 1 - \max\{m_a, m_b, m_c\}, \\ \quad \text{if } \max\{m_a, m_b, m_c\} > -\min\{m_a, m_b, m_c\}; \\ \min\{m_a, m_b, m_c\} - 1, \\ \quad \text{if } \max\{m_a, m_b, m_c\} \le -\min\{m_a, m_b, m_c\}. \end{cases}$$

(8)

TABLE III
Setup specifications

| Module inductor | $L_{mdl}$ | 10 μH (744711015, Würth Electronics) |
|---|---|---|
| Module capacitor | $C_{mdl}$ | 1.5 mF/25 V |
| Module battery | | Lithium Polymer (5.2 Ah, 16.4 V) |
| Module transistors | | 300 A/30 V (IPT004N03L, Infineon) |
| DC-link inductor | $L$ | 15 μH × 2 (7443641500, Würth Electronics) |
| DC-link capacitor | $C$ | 30 μF × 2 (C4AQQBW5300A3MJ, KEMET) |
| Frontend IGBT | | 650 V/100 A (FS75R07W2E3, Infineon) |
| Backend carrier frequency | $f_{mdl}$ | 5 kHz (effectively 40 kHz at $v_{dc1}$) |
| Frontend carrier frequency | $f_{inv}$ | 10 kHz (effectively 3.3 kHz per switch) |

TABLE IV
Load specifications

| **Induction motor** | | |
|---|---|---|
| Nominal power | $P_{motor}$ | 250 W |
| Nominal voltage | $V$ | 230 V |
| Nominal speed | $f_{motor}$ | 1800 rpm |
| **Passive load** | | |
| Total power | $P_{load}$ | 2.7 kW |
| Modulation index | $m$ | 0.95 |
| Load resistance | $R_{load}$ | 2.2 Ω |
| Load inductance | $L_{load}$ | 100 μH |
| Load frequency | $f_{load}$ | 50 Hz |

so that $1/3$ of the switches remain untoggled. Even compared to DPWM, the proposed solution still halves the switching actions.

The detailed simulation settings are listed in TABLE II. In the conventional setup, the dc link is fixed at 640 V and the main inverter is kept the same as that of the proposed system (FF300R12ME4, Infineon). Across all cases, we fix the peak output current to 200 A but vary the modulation index and the power factor. Low modulation indices emulate smaller back emf under low speeds, whereas the combination of higher modulation indices and low power factors (cos $\varphi$) reflect field-weakening at high speeds.

Fig. 3 details the converter losses. Despite the additional loss in the FETs, the proposed topology notably reduces the overall loss under all conditions thanks to the fewer switching instances in the frontend inverter. The relative ratio of the IGBT switching losses for the three evaluated cases is approximately 1:2:3 as predicted. The IGBTs' conduction losses are identical across all cases because of the matched output current; the FET loss in the reconfigurable battery decreases at lower power factors due to the cancellation effect at the dc link.

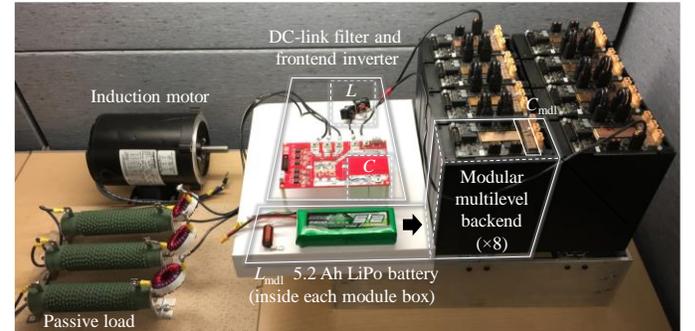

Fig. 7. Experimental setup. $L$ = 30 μH, $C$ = 60 μF, and $L_{mdl}$ = 10 μH.

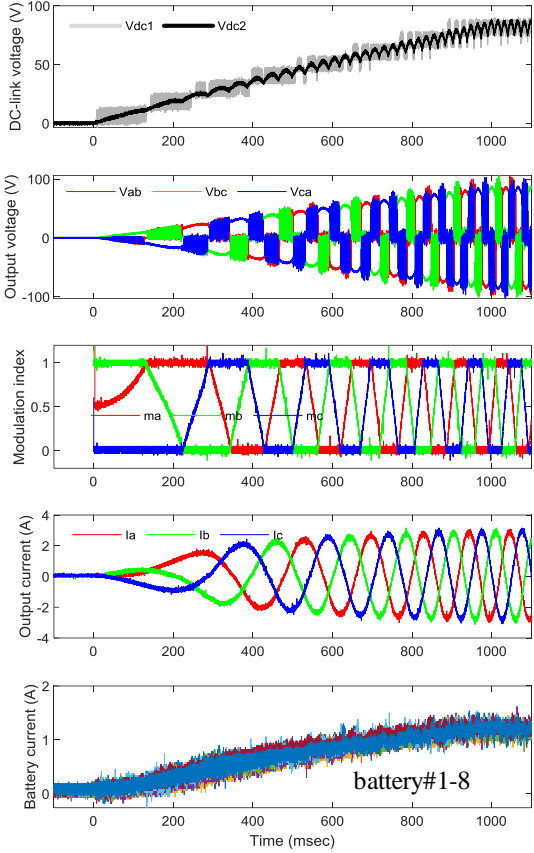

Fig. 8. Individual components and output during the start-up of a motor representing a wider speed range.

We select FF300R12ME4 (IGBT) and IPT007N06N (FET) based on the availability of accurate models and the matched cost. The relative benefit of the proposed solution depends on the choice of the frontend IGBT. There are several motor-drive-specific features favoring the proposed solution: (1) at low speeds and thus smaller back emf, the frontend IGBTs of the solution experience significantly less switching stress as is evidenced by Fig. 3(a)–(c); (2) at high speeds and thus lower power factor due to field-weakening, a large share of the currents cancel at the dc link so that they do not reach the battery backend, producing small conduction loss at the FETs; and (3) the FET loss is mostly the conduction loss, which can be reduced and easily scaled for higher current loads by paralleling more silicon, whereas the IGBTs' switching loss does not decrease in the same manner.

We further use passive loads to evaluate the THDs of the three configurations. For each phase $R_{load}$ = 1.75 Ω, $L_{load}$ = 200 µH. The results are shown in Fig. 5. The proposed converter maintains a low THD (5.3%) across all modulation indices whereas both DPWM and SVPWM suffer higher distortion at lower modulation indices. For DPWM, THD can be almost four times higher at low modulation indices, although both the switching and the total loss are substantially larger.

### D. Design of the L-C Filter

The size of the batteries and the wiring in EV drive systems entails nonnegligible stray inductances in the wiring (Fig. 7). As such, there must be a (small) dc-link capacitor placed near the frontend inverter for the switching transients. An inductor $L$ is consequently inserted to suppress the surge current, forming an L-C filter between $v_{dc1}$ and $v_{dc2}$.

The L-C filter introduces phase lag and amplitude change into $v_{dc2}$, which can be considered by the control. Fig. 6 quantifies these effects, where the frontend inverter is modeled by $R_{eq}$. At 136 kW and thus $R_{eq}$ = 3 Ω, $v_{dc2}$ is delayed by 3.9° and amplified by 7.6% when $v_{dc1}$ pulsates at 1 kHz.

## VI. EXPERIMENTS

We implemented the adaptive dc battery with eight cascaded modules. Each module contains a FET half-bridge (Infineon IPT004N03L), a four-cell LiFePO$_4$ battery (16.4 V, 5.2 Ah), and an L-C filter with $L_{mdl}$ = 10 µH (Würth Electronics 744711015), $C_{mdl}$ = 1.5 mF (Murata KRM55WR72A) to smoothen the battery current. The frontend inverter used three IGBT half-bridges (Infineon FS75R07W2E3). Between the backend reconfigurable battery and the frontend inverter, we implemented a dc-link filter with $L$ = 30 µH (Würth Electronics 7443641500) and $C$ = 60 µF (Kemet C4AQQBBW5300A3MJ), which resonates at 3.75 kHz. To avoid resonances with the filter, the frontend motor inverter was modulated with a 10-kHz carrier. Under the proposed modulation scheme, each transistor switches at 3.33 kHz on average; whereas the backend modules switch at 5 kHz each and thus effectively 40 kHz at $v_{dc1}$. Both the frontend inverter and the backend reconfigurable battery are controlled by an FPGA (sbRIO 9627 with Zynq 7020, National Instruments). The converter setup is detailed in TABLE III and shown in Fig. 7. All passive components and backend electronics exceed the batteries' current rating (Fig. 9), yet they have comparably negligible size and cost (see $L$, $C$, $L_{mdl}$, $C_{mdl}$, and the surface-mounted FETs atop the backend modules in Fig. 7).

We tested the setup with a three-phase induction motor. Fig. 8 shows the spin-up of the motor (TABLE IV). The motor voltage ramps up to some 90 V pk/65 V rms with increasing speed. The backend battery not only follows the ideal voltage but also provides fine-shaping for the ac output of the inverter frontend. Accordingly, the battery voltage profile of $v_{dc1}$ clearly shows both contributions and very fine switching modulation due to the fast FETs. The high-quality waveform of $v_{dc1}$ justifies the use of just a small L-C filter, which also enables fast response dynamics [27]. The inverter frontend operates practically merely as a commutator in the over-modulation range with close to minimum switching. In the upper and lower plateaus, the adaptive battery is providing all output fine-shaping, while the specific frontend inverter phase stays static, avoiding lossy switching of the IGBTs, particularly during high-current phases. Most prominently, the adaptive battery backend contributes fast output content, which emerges on the battery side as sixth harmonic of the output frequency associated with rectification and three phases. Despite the seemingly rough switching of the frontend inverter, the current is smooth with sinusoidal shape (Fig. 9). The system avoids low-frequency ripple load on the individual battery modules, which is known to be a major problem of many other reconfigurable battery setups [66], [67], [79].

For a closer analysis, we connected a three-phase inductive load where each phase is formed by 2.2 Ω and 100 µH in series to avoid any distortion from the motor for a fair analysis of the output. In contrast to the typical three levels and intensive switching noise of conventional inverters that makes the visual identification of the sinusoidal voltage content practically impossible, the voltage of the proposed system obviously has multiple levels (black line, Panel 2 for phase-to-phase and Panel 3 for phase-to-neutral). The modulation indices (fourth panel) clearly shows the contribution of the adaptive battery backend (black line), which provides the sinusoidal caps and the fine shaping, and the frontend inverter (colors), which are maxed out most of the time. As demonstrated in Fig. 9, the dc-link current is relatively constant with only minor modulation (red line, first panel). The battery backend spreads out particularly the low-frequency content of this residual ripple across all battery modules (bottom panel of Fig. 9) as it was suggested as dominant in loss and battery ageing potential [79], [97]. The comparison of the phase voltage $v_a$, phase current $i_a$, and modulation reference $m_a$ underlines that (1) only a third of the frontend inverter transistors are switching at each time; and (2) the frontend inverter transistors exclusively switch during periods with low currents. The other phases behave equivalently but shifted by 120°. As long as the power factor $\cos\varphi > 0.5$, the current peaks can be shaped solely by the backend battery, which enables small switching ripple and distortion and still low switching effort.

In addition to output shaping, the load in Fig. 9 was intentionally shifted through carrier offsetting to demonstrate active power control on the module level. Load is shifted from the third battery module to the first, which leads to different means of the individual currents. The head room for shifting load increases at lower modulation indices. The approximately 5% balancing current (at 95% modulation index) notably exceeds typical balancing currents in conventional battery balancing systems for charge balancing (e.g., typically ≤ 200 mA for 500 A load current, thus 0.04%), and might also be usable to counteract local thermal hotspots through power redistribution.

## VII. Conclusion

The presented distribution of output modulation across a rapidly reconfigurable dc battery and a drive inverter demonstrated lower overall switching loss, superior distortion and low d$v$/d$t$ stress. Thus, it has great potential for reducing motor insulation stress, bearing currents, and switching loss as well as increasing the output quality and round-trip efficiency. Previous solutions, typically with dc/dc converters, have various drawbacks including limited operating range, no support in active battery management, stability challenges, as well as marginal efficiency gain that can barely justify the additional high-voltage high-power transistors and large magnetic components with their substantial cost.

Instead, this paper proposes and comprehensively exploits a modular adaptively reconfigurable multilevel dc battery to modulate the input voltage of a subsequent six-switch motor inverter—a technology that is also studied by the vehicle industry, though for quasi-static voltage adjustment and battery management. The modular adaptive dc battery allows the use of latest high-performance low-voltage FETs and thus surrogates hard switching of the main inverter at a negligible switching effort. Despite the loss in the FETs, the proposed solution notably reduces the overall loss under various modulation indices and power factors for a wide range of operating conditions. Additionally, the adaptively reconfigurable dc battery offers active balancing of capacitance, power, as well as ageing and eliminates the high vulnerability from weak cells in conventional hard-wired batteries, while further turning the drive train to a normally-off safe low-voltage system. The advantages of the proposed solution are quantified by accurate SPICE simulations and verified by experiments.

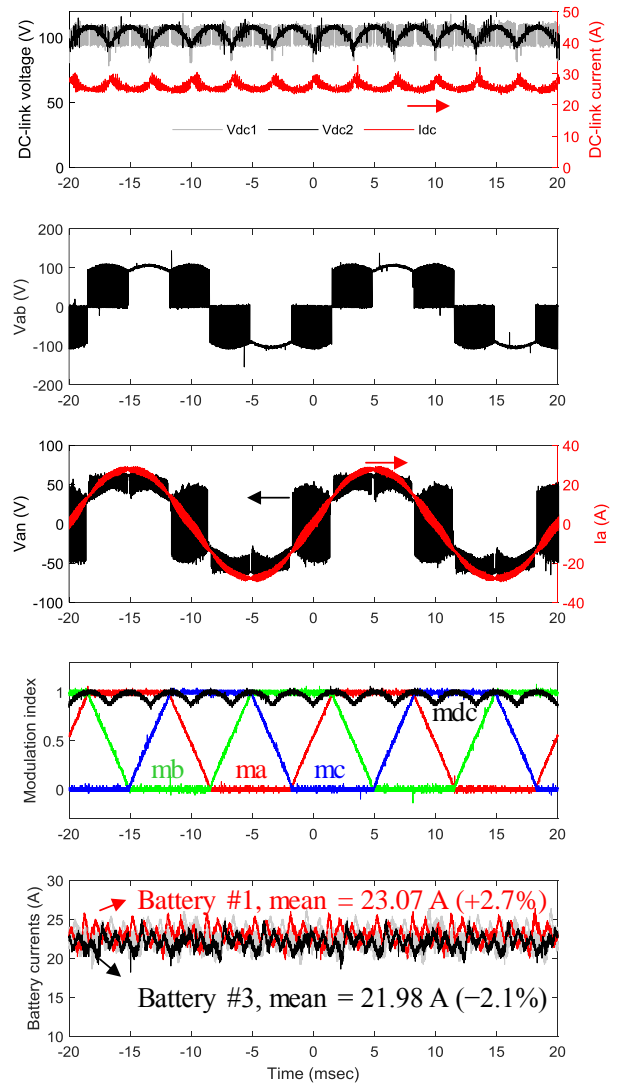

Fig. 9. Waveforms under 3 kW three-phase passive load at 95% modulation index (i.e., 95% × 1.13 = 107% utilization of the dc-link voltage. Balancing is active between batteries #1 and #3. The time is in milliseconds.

## Reference

[1] B. A. Welchko and J. M. Nagashima, "A comparative evaluation of motor drive topologies for low-voltage, high-power EV/HEV propulsion systems," in *IEEE International Symposium on Industrial Electronics*, 2003, vol. I, no. 1, pp. 379–384. doi: 10.1109/ISIE.2003.1267278.

[2] J. O. Estima and A. J. Marques Cardoso, "Efficiency analysis of drive train topologies applied to electric/hybrid vehicles," *IEEE Trans. Veh. Technol.*, vol. 61, no. 3, pp. 1021–1031, Mar. 2012, doi: 10.1109/TVT.2012.2186993.

[3] J. Reimers, L. Dorn-Gomba, C. Mak, and A. Emadi, "Automotive Traction Inverters: Current Status and Future Trends," *IEEE Trans. Veh. Technol.*, vol. 68, no. 4, pp. 3337–3350, 2019, doi: 10.1109/TVT.2019.2897899.

[4] K. Rahman *et al.*, "The Voltec 4ET50 Electric Drive System," *SAE Int. J. Engines*, vol. 4, no. 1, pp. 323–337, Apr. 2011, doi: 10.4271/2011-01-0355.
[5] C. Jung, "Power Up with 800-V Systems: The benefits of upgrading voltage power for battery-electric passenger vehicles," *IEEE Electrif. Mag.*, vol. 5, no. 1, pp. 53–58, Mar. 2017, doi: 10.1109/MELE.2016.2644560.
[6] C. Y. Yu, J. Tamura, and R. D. Lorenz, "Optimum dc bus voltage analysis and calculation method for inverters/motors with variable dc bus voltage," *IEEE Trans. Ind. Appl.*, vol. 49, no. 6, pp. 2619–2627, Nov. 2013, doi: 10.1109/TIA.2013.2265873.
[7] P. N. Tekwani and P. V. Manilal, "Novel approach employing buck-boost converter as DC-link modulator and inverter as AC-chopper for induction motor drive applications: An alternative to conventional AC-DC-AC scheme," in *IEEE International Symposium on Industrial Electronics*, Jun. 2017, pp. 793–800. doi: 10.1109/ISIE.2017.8001347.
[8] D. C. Rus, N. S. Preda, I. I. Incze, M. Imecs, and C. Szabó, "Comparative analysis of PWM techniques: Simulation and DSP implementation," 2010. doi: 10.1109/AQTR.2010.5520764.
[9] P. Ide, N. Froehleke, and H. Grotstollen, "Comparison of selected 3-phase switched mode rectifiers," 1997. doi: 10.1109/intlec.1997.646061.
[10] A. Chen, "Electric vehicles mean first responders have to deal with battery fires," *The Verge*, Jul. 2018.
[11] N. Gordon-Bloomfield, "Facing a Wrecked Electric Vehicle, What Must EMS Staff Know?," *Green Car Reports*, 2010.
[12] A. Marellapudi, M. J. Mauger, P. Kandula, and D. Divan, "Intrinsically-Safe Modular Power Converters for Electric Transportation," in *2020 IEEE Transportation Electrification Conference & Expo (ITEC)*, Jun. 2020, pp. 1–6. doi: 10.1109/ITEC48692.2020.9161533.
[13] R. T. Long Jr. *et al.*, "Best Practices for Emergency Response to Incidents Involving Electric Vehicles Battery Hazards : A Report on Full-Scale Testing Results," *J. Chem. Inf. Model.*, 2013.
[14] Jesse Roman, "Stranded Energy," *The magazine of the National Fire Protection Association*, 2020.
[15] J. O. Estima and A. J. Marques Cardoso, "Performance analysis of a PMSM drive for hybrid electric vehicles," in *The XIX International Conference on Electrical Machines - ICEM 2010*, Sep. 2010, pp. 1–6. doi: 10.1109/ICELMACH.2010.5608285.
[16] S. M. Lukic, J. Cao, R. C. Bansal, F. Rodriguez, and A. Emadi, "Energy storage systems for automotive applications," *IEEE Trans. Ind. Electron.*, vol. 55, no. 6, pp. 2258–2267, Jun. 2008, doi: 10.1109/TIE.2008.918390.
[17] S. Rivera, S. Kouro, S. Vazquez, S. M. Goetz, R. Lizana, and E. Romero-Cadaval, "Electric Vehicle Charging Infrastructure: From Grid to Battery," *IEEE Ind. Electron. Mag.*, vol. 15, no. 2, pp. 37–51, 2021, doi: 10.1109/MIE.2020.3039039.
[18] D. Herke and D. Spesser, "Method for charging a direct current traction battery at a direct current charging pillar," US Patent 10,471,837, 2019
[19] S. Götz and T. Lütje, "Inverter for an electric automobile," US Patent 10,505,439, 2019
[20] L. Yang, Y. Luo, R. S. K. Moorthy, D. Rahman, W. Yu, and I. Husain, "Design and Test of a Planarized High Power Density 100 kW SiC Traction Inverter with 1kV DC-Link," 2018. doi: 10.1109/ECCE.2018.8557467.
[21] I. Aghabali, J. Bauman, and A. Emadi, "Analysis of Auxiliary Power Unit and Charging for an 800V Electric Vehicle," in *2019 IEEE Transportation Electrification Conference and Expo (ITEC)*, Jun. 2019, pp. 1–6. doi: 10.1109/ITEC.2019.8790562.
[22] J. S. Hsu, C. W. Ayers, and C. L. Coomer, "Report on Toyota/Prius motor design and manufacturing assessment," *Oak Ridge Natl. Lab.*, p. 14, 2010.
[23] Z. Du, B. Ozpineci, L. M. Tolbert, and J. N. Chiasson, "DC-AC cascaded H-bridge multilevel boost inverter with no inductors for electric/hybrid electric vehicle applications," *IEEE Trans. Ind. Appl.*, vol. 45, no. 3, pp. 963–970, 2009, doi: 10.1109/TIA.2009.2018978.
[24] M. Shen, F. Z. Peng, and L. M. Tolbert, "Multi-level DC/DC power conversion system with multiple DC sources," in *PESC Record - IEEE Annual Power Electronics Specialists Conference*, 2007, vol. 23, no. 1, pp. 2008–2014. doi: 10.1109/PESC.2007.4342313.
[25] C. Klumpner, "A new two-stage voltage source inverter with modulated DC-link voltage and reduced switching losses," in *IECON Proceedings (Industrial Electronics Conference)*, Nov. 2006, pp. 2208–2213. doi: 10.1109/IECON.2006.347904.
[26] P. Liutanakul, A. Awan, S. Pierfederici, B. Nahid-Mobarakeh, and F. Meibody-Tabar, "Linear Stabilization of a DC Bus Supplying a Constant Power Load: A General Design Approach," *IEEE Trans. Power Electron.*, vol. 25, no. 2, pp. 475–488, 2010, doi: 10.1109/TPEL.2009.2025274.
[27] J. Kacetl, J. Fang, T. Kacetl, N. Tashakor, and S. Goetz, "Design and Analysis of Modular Multilevel Reconfigurable Battery Converters for Variable Bus Voltage Powertrains," *IEEE Trans. Power Electron.*, 2022, doi: 10.1109/TPEL.2022.3179285.
[28] J. Engelhardt, T. Gabderakhmanova, G. Rohde, and M. Marinelli, "Reconfigurable Stationary Battery with Adaptive Cell Switching for Electric Vehicle Fast-Charging," in *2020 55th International Universities Power Engineering Conference (UPEC)*, Sep. 2020, pp. 1–6. doi: 10.1109/UPEC49904.2020.9209774.
[29] S. Goetz, R. Bauer, H. Schoeffler, and J. Mittnacht, "Battery system with battery control," US Patent App. 15/095,409, 2016
[30] M. Jaensch, J. Kacetl, T. Kacetl, and S. Götz, "Vehicle having an energy storage element," US 11,088,550, 2021
[31] T. Morstyn, M. Momayyezan, B. Hredzak, and V. G. Agelidis, "Distributed Control for State-of-Charge Balancing Between the Modules of a Reconfigurable Battery Energy Storage System," *IEEE Trans. Power Electron.*, vol. 31, no. 11, pp. 7986–7995, Nov. 2016, doi: 10.1109/TPEL.2015.2513777.
[32] P. Wacker, J. Adermann, B. Danquah, and M. Lienkamp, "Efficiency determination of active battery switching technology on roller dynamometer," in *2017 Twelfth International Conference on Ecological Vehicles and Renewable Energies (EVER)*, Apr. 2017, pp. 1–7. doi: 10.1109/EVER.2017.7935866.
[33] F. Chen, W. Qiao, and L. Qu, "A modular and reconfigurable battery system," in *2017 IEEE Applied Power Electronics Conference and Exposition (APEC)*, Mar. 2017, pp. 2131–2135. doi: 10.1109/APEC.2017.7930993.
[34] F. Ji, L. Liao, T. Wu, C. Chang, and M. Wang, "Self-reconfiguration batteries with stable voltage during the full cycle without the DC-DC converter," *J. Energy Storage*, vol. 28, p. 101213, Apr. 2020, doi: 10.1016/j.est.2020.101213.
[35] J. Cao, J. Wang, S. Zhang, and M. Li, "A dynamically reconfigurable system based on workflow and service agents," *Eng. Appl. Artif. Intell.*, vol. 17, no. 7, pp. 771–782, Oct. 2004, doi: 10.1016/j.engappai.2004.08.030.
[36] F. Chen, H. Wang, W. Qiao, and L. Qu, "A grid-tied reconfigurable battery storage system," in *2018 IEEE Applied Power Electronics Conference and Exposition (APEC)*, Mar. 2018, pp. 645–652. doi: 10.1109/APEC.2018.8341080.
[37] H. Kim and K. G. Shin, "Dynamically reconfigurable framework for a large-scale battery system," US Patent 8,330,420, 2012
[38] S. Goetz, R. Bauer, H. Schoeffler, and J. Mittnacht, "Battery system," US Patent 10,211,485, 2019
[39] Y. Cheng, C. Lu, and C. Liu, "Battery management system and vehicle," WO 2021 227 589, 2021
[40] M. Slepchenkov and R. Naderi, "Module-based energy systems having converter-source modules and methods related thereto," US 10,807,481, 2020
[41] M. Slepchenkov and R. Naderi, "Module-based energy systems capable of cascaded and interconnected configurations, and methods related thereto," US 11,135,923, 2019
[42] S. Goetz, J. Kacetl, T. Kacetl, and M. Jaensch, "Electric energy storage system," DE 10 2018 109 921, 2019
[43] K. Kristensen, "Method and apparatus for creating a dynamically reconfigurable energy storage device," US 10,305,298, 2019
[44] J. V. Muenzel and D. Crowley, "Battery system including circuit module for selectively connecting a plurality of battery cell units," US 10,573,935, Feb. 2020
[45] M. Braun and H. Rapp, "Energy storage device having a dc voltage supply circuit and method for providing a dc voltage from an energy storage device," EP 3 014 725, 2017
[46] A. Xie, "Modular inverters with batteries as energy storage for three-phase electric motors," DE 10 2018 003 642, 2018
[47] A. Xie, "Modular energy storage with parallel battery strings," DE 10 2018 007 919, 2020
[48] N. Tashakor, J. Kacetl, J. Fang, Z. Li, and S. Goetz, "Dual-Port Dynamically Reconfigurable Battery with Semi-Controlled and Fully-Controlled Outputs." arXiv, 2022. doi: 10.48550/ARXIV.2206.01435.
[49] N. Tashakor, B. Arabsalmanabadi, F. Naseri, and S. Goetz, "Low-Cost Parameter Estimation Approach for Modular Converters and Reconfigurable Battery Systems Using Dual Kalman Filter," *IEEE Trans. Power Electron.*, vol. 37, no. 6, pp. 6323–6334, 2022, doi: 10.1109/TPEL.2021.3137879.
[50] S. M. Goetz, Z. Li, X. Liang, C. Zhang, S. M. Lukic, and A. V. Peterchev, "Control of Modular Multilevel Converter with Parallel Connectivity-Application to Battery Systems," *IEEE Trans. Power Electron.*, vol. 32, no. 11, pp. 8381–8392, 2017, doi: 10.1109/TPEL.2016.2645884.

[51] R. K. Williams, M. N. Darwish, R. A. Blanchard, R. Siemieniec, P. Rutter, and Y. Kawaguchi, "The Trench Power MOSFET: Part I—History, Technology, and Prospects," *IEEE Trans. Electron Devices*, vol. 64, no. 3, pp. 674–691, Mar. 2017, doi: 10.1109/TED.2017.2653239.

[52] R. K. Williams, M. N. Darwish, R. A. Blanchard, R. Siemieniec, P. Rutter, and Y. Kawaguchi, "The Trench Power MOSFET—Part II: Application Specific VDMOS, LDMOS, Packaging, and Reliability," *IEEE Trans. Electron Devices*, vol. 64, no. 3, pp. 692–712, Mar. 2017, doi: 10.1109/TED.2017.2655149.

[53] Z. Li, A. Yang, G. Chen, Z. Zeng, A. V Peterchev, and S. M. Goetz, "A High-Frequency Pulsating DC-Link for Electric Vehicle Drives with Reduced Losses," in *IECON 2021 – 47th Annual Conference of the IEEE Industrial Electronics Society*, 2021, pp. 1–6. doi: 10.1109/IECON48115.2021.9589040.

[54] N. R. Raju, "A DC link-modulated three-phase converter," in *Conference Record of the 2001 IEEE Industry Applications Conference. 36th IAS Annual Meeting (Cat. No.01CH37248)*, 2002, vol. 4, no. C, pp. 2181–2185. doi: 10.1109/IAS.2001.955928.

[55] N. R. Raju, A. Daneshpooy, and J. Schwartzenberg, "Harmonic cancellation for a twelve-pulse rectifier using DC bus modulation," in *Conference Record of the 2002 IEEE Industry Applications Conference. 37th IAS Annual Meeting (Cat. No.02CH37344)*, 2003, vol. 4, pp. 2526–2529. doi: 10.1109/IAS.2002.1042801.

[56] H. P. Felsl, F.-J. Niedemostheide, and H.-J. Schulze, "IGBT field-stop design for good short circuit ruggedness and a better trade-off with respect to static and dynamic switching characteristics," in *2017 29th International Symposium on Power Semiconductor Devices and IC's (ISPSD)*, 2017, pp. 143–146. doi: 10.23919/ISPSD.2017.7988931.

[57] F. Wolter, W. Roesner, M. Cotorogea, T. Geinzer, M. Seider-Schmidt, and K.-H. Wang, "Multi-dimensional trade-off considerations of the 750V micro pattern trench IGBT for electric drive train applications," in *2015 IEEE 27th International Symposium on Power Semiconductor Devices & IC's (ISPSD)*, 2015, pp. 105–108. doi: 10.1109/ISPSD.2015.7123400.

[58] J. Hu, M. Bobde, H. Yilmaz, and A. Bhalla, "Trench shielded planar gate IGBT (TSPG-IGBT) for low loss and robust short-circuit capability," in *2013 25th International Symposium on Power Semiconductor Devices & IC's (ISPSD)*, 2013, pp. 25–28. doi: 10.1109/ISPSD.2013.6694390.

[59] M. Tsirinomeny and A. Rufer, "Configurable Modular Multilevel Converter (CMMC) for flexible EV," in *2015 17th European Conference on Power Electronics and Applications (EPE'15 ECCE-Europe)*, Sep. 2015, no. Cmmc, pp. 1–10. doi: 10.1109/EPE.2015.7309229.

[60] Z. Zheng, K. Wang, L. Xu, and Y. Li, "A Hybrid Cascaded Multilevel Converter for Battery Energy Management Applied in Electric Vehicles," *IEEE Trans. Power Electron.*, vol. 29, no. 7, pp. 3537–3546, Jul. 2014, doi: 10.1109/TPEL.2013.2279185.

[61] Fa Chen, Wei Qiao, and Liyan Qu, "A scalable and reconfigurable battery system," in *2017 IEEE 8th International Symposium on Power Electronics for Distributed Generation Systems (PEDG)*, Apr. 2017, pp. 1–6. doi: 10.1109/PEDG.2017.7972549.

[62] F. Chang, O. Ilina, M. Lienkamp, and L. Voss, "Improving the Overall Efficiency of Automotive Inverters Using a Multilevel Converter Composed of Low Voltage Si mosfets," *IEEE Trans. Power Electron.*, vol. 34, no. 4, pp. 3586–3602, Apr. 2019, doi: 10.1109/TPEL.2018.2854756.

[63] F. Khoucha, S. M. Lagoun, K. Marouani, A. Kheloui, and M. El Hachemi Benbouzid, "Hybrid Cascaded H-Bridge Multilevel-Inverter Induction-Motor-Drive Direct Torque Control for Automotive Applications," *IEEE Trans. Ind. Electron.*, vol. 57, no. 3, pp. 892–899, Mar. 2010, doi: 10.1109/TIE.2009.2037105.

[64] S. M. Goetz, Z. Li, A. V Peterchev, X. Liang, C. Zhang, and S. M. Lukic, "Sensorless scheduling of the modular multilevel series-parallel converter: enabling a flexible, efficient, modular battery," in *2016 IEEE Applied Power Electronics Conference and Exposition (APEC)*, 2016, pp. 2349–2354. doi: 10.1109/APEC.2016.7468193.

[65] Z. Li, R. Lizana, A. V Peterchev, and S. M. Goetz, "Predictive control of modular multilevel series/parallel converter for battery systems," in *2017 IEEE Energy Conversion Congress and Exposition (ECCE)*, 2017, pp. 5685–5691. doi: 10.1109/ECCE.2017.8096945.

[66] Z. Li, R. Lizana, A. V Peterchev, S. Member, and S. M. Goetz, "Ripple Current Suppression Methods for Star- Configured Modular Multilevel Converters".

[67] Z. Li, R. Lizana, Z. Yu, S. Sha, A. V. Peterchev, and S. M. Goetz, "Modulation and Control of Series/Parallel Module for Ripple-Current Reduction in Star-Configured Split-Battery Applications," *IEEE Trans. Power Electron.*, vol. 35, no. 12, pp. 12977–12987, Dec. 2020, doi: 10.1109/TPEL.2020.2996542.

[68] B. Arabsalmanabadi, N. Tashakor, Y. Zhang, K. Al-Haddad, and S. Goetz, "Parameter Estimation of Batteries in MMCs with Parallel Connectivity using PSO," in *IECON 2021 – 47th Annual Conference of the IEEE Industrial Electronics Society*, 2021, pp. 1–6. doi: 10.1109/IECON48115.2021.9589293.

[69] J. B. Quinn, T. Waldmann, K. Richter, M. Kasper, and M. Wohlfahrt-Mehrens, "Energy Density of Cylindrical Li-Ion Cells: A Comparison of Commercial 18650 to the 21700 Cells," *J. Electrochem. Soc.*, vol. 165, no. 14, pp. A3284–A3291, Oct. 2018, doi: 10.1149/2.0281814jes.

[70] Y. Ding, Z. P. Cano, A. Yu, J. Lu, and Z. Chen, "Automotive Li-Ion Batteries: Current Status and Future Perspectives," *Electrochem. Energy Rev.*, vol. 2, no. 1, pp. 1–28, Mar. 2019, doi: 10.1007/s41918-018-0022-z.

[71] Z. Li, R. Lizana, S. M. Lukic, A. V. Peterchev, and S. M. Goetz, "Current Injection Methods for Ripple-Current Suppression in Delta-Configured Split-Battery Energy Storage," *IEEE Trans. Power Electron.*, vol. 34, no. 8, pp. 7411–7421, 2019, doi: 10.1109/TPEL.2018.2879613.

[72] K. Qian, C. Zhou, Y. Yuan, and M. Allan, "Temperature effect on electric vehicle battery cycle life in Vehicle-to-grid applications," 2010.

[73] J.-M. Tarascon, A. S. Gozdz, C. Schmutz, F. Shokoohi, and P. C. Warren, "Performance of Bellcore's plastic rechargeable Li-ion batteries," *Solid State Ionics*, vol. 86–88, no. PART 1, pp. 49–54, Jul. 1996, doi: 10.1016/0167-2738(96)00330-X.

[74] A. J. Ruddell *et al.*, "Analysis of battery current microcycles in autonomous renewable energy systems," *J. Power Sources*, vol. 112, no. 2, pp. 531–546, Nov. 2002, doi: 10.1016/S0378-7753(02)00457-3.

[75] T. Note and B. Continuity, "Effects of AC Ripple Current on VRLA Battery Life," *A Tech. Note from Expert Businnes-Critical Contin.*, p. 8, 2013.

[76] C. Ropeter, H. Wenzl, H. Beck, and E. A. Wehrmann, "The Impact of Microcycles on Batteries in Different Applications," *Proc. 18th Electr. Veh. Symp. (EVS18), Berlin*, pp. 1–12, 2001, doi: 10.1.1.580.4107.

[77] R. F. Nelson and M. A. Kepros, "AC ripple effects on VRLA batteries in float applications," *Batter. Conf. Appl. Adv. 1999. Fourteenth Annu.*, pp. 281–289, 2008, doi: 10.1109/bcaa.1999.796005.

[78] A. I. Harrison, "Batteries and AC phenomena in UPS systems: the battery point of view," in *Conference Proceedings., Eleventh International Telecommunications Energy Conference*, 1989, pp. 12.5/1-12.5/6. doi: 10.1109/INTLEC.1989.88290.

[79] T. Kacetl, J. Kacetl, J. Fang, M. Jaensch, and S. Goetz, "Degradation-Reducing Control for Dynamically Reconfigurable Batteries." arXiv, 2022. doi: 10.48550/ARXIV.2202.11757.

[80] C. Klumpner and T. R. Zargar, "Hybrid inverter arrangements to facilitate reduced switching losses of the main inverter," in *IECON Proceedings (Industrial Electronics Conference)*, Oct. 2016, no. 1, pp. 3425–3430. doi: 10.1109/IECON.2016.7793802.

[81] U. R. Prasanna and A. K. Rathore, "Dual Three-Pulse Modulation-Based High-Frequency Pulsating DC Link Two-Stage Three-Phase Inverter for Electric/Hybrid/Fuel Cell Vehicles Applications," *IEEE J. Emerg. Sel. Top. Power Electron.*, vol. 2, no. 3, pp. 477–486, Sep. 2014, doi: 10.1109/jestpe.2014.2304472.

[82] J. Fang and S. M. Goetz, "Symmetries in Power Electronics," in *2021 IEEE Energy Conversion Congress and Exposition (ECCE)*, 2021, pp. 1261–1268. doi: 10.1109/ECCE47101.2021.9594928.

[83] L. M. Tolbert, F. Z. Peng, and T. G. Habetler, "Multilevel converters for large electric drives," *IEEE Trans. Ind. Appl.*, vol. 35, no. 1, pp. 36–44, 1999, doi: 10.1109/28.740843.

[84] J. I. Y. Ota, T. Sato, and H. Akagi, "Enhancement of Performance, Availability, and Flexibility of a Battery Energy Storage System Based on a Modular Multilevel Cascaded Converter (MMCC-SSBC)," *IEEE Trans. Power Electron.*, vol. 31, no. 4, pp. 2791–2799, Apr. 2016, doi: 10.1109/TPEL.2015.2450757.

[85] L. Baruschka and A. Mertens, "Comparison of cascaded H-bridge and modular multilevel converters for BESS application," in *IEEE Energy Conversion Congress and Exposition: Energy Conversion Innovation for a Clean Energy Future, ECCE 2011, Proceedings*, Sep. 2011, pp. 909–916. doi: 10.1109/ECCE.2011.6063868.

[86] J. Svensson, "Improved power system stability and reliability using innovative energy storage technologies," in *8th IEE International Conference on AC and DC Power Transmission (ACDC 2006)*, 2006, vol. 2006, pp. 220–224. doi: 10.1049/cp:20060045.

[87] Y. Iijima *et al.*, "Development and field experiences of NAS battery inverter for power stabilization of a 51 MW wind farm," *2010 Int. Power Electron. Conf. - ECCE Asia -, IPEC 2010*, pp. 1837–1841, Jun. 2010, doi: 10.1109/IPEC.2010.5543520.

[88] H. Li, Y. Iijima, and N. Kawakami, "Development of power conditioning system (PCS) for battery energy storage systems," in *2013 IEEE ECCE*

*Asia Downunder - 5th IEEE Annual International Energy Conversion Congress and Exhibition, IEEE ECCE Asia 2013*, Jun. 2013, pp. 1295–1299. doi: 10.1109/ECCE-Asia.2013.6579276.

[89] L. H. Walker, "10-MW GTO Converter for Battery Peaking Service," *IEEE Trans. Ind. Appl.*, vol. 26, no. 1, pp. 63–72, 1990, doi: 10.1109/28.52675.

[90] N. W. Miller, R. S. Zrebiec, G. Hunt, and R. W. Deimerico, "Design and commissioning of a 5 MVA, 2.5 MWh battery energy storage system," in *Proceedings of 1996 Transmission and Distribution Conference and Exposition*, 2002, no. M, pp. 339–345. doi: 10.1109/TDC.1996.545957.

[91] N. Tashakor, M. Kilictas, E. Bagheri, and S. Goetz, "Modular Multilevel Converter With Sensorless Diode-Clamped Balancing Through Level-Adjusted Phase-Shifted Modulation," *IEEE Trans. Power Electron.*, vol. 36, no. 7, pp. 7725–7735, 2021, doi: 10.1109/TPEL.2020.3041599.

[92] Z. Li, R. Lizana, A. V Peterchev, and S. M. Goetz, "Distributed balancing control for modular multilevel series/parallel converter with capability of sensorless operation," in *2017 IEEE Energy Conversion Congress and Exposition (ECCE)*, 2017, pp. 1787–1793. doi: 10.1109/ECCE.2017.8096011.

[93] Z. Li, R. Lizana, S. Sha, Z. Yu, A. V. Peterchev, and S. M. Goetz, "Module implementation and modulation strategy for sensorless balancing in modular multilevel converters," *IEEE Trans. Power Electron.*, vol. 34, no. 9, pp. 8405–8416, 2019, doi: 10.1109/TPEL.2018.2886147.

[94] J. Asakura and H. Akagi, "State-of-Charge (SOC)-Balancing Control of a Battery Energy Storage System Based on a Cascade PWM Converter," *IEEE Trans. Power Electron.*, vol. 24, no. 6, pp. 1628–1636, 2009, doi: 10.1109/TPEL.2009.2014868.

[95] M. Vasiladiotis and A. Rufer, "Analysis and Control of Modular Multilevel Converters With Integrated Battery Energy Storage," *IEEE Transactions on Power Electronics*, vol. 30, no. 1. pp. 163–175, 2015. doi: 10.1109/TPEL.2014.2303297.

[96] E. R. C. da Silva and M. E. Elbuluk, *Fundamentals of Power Electronics*, vol. 59. Boston, MA: Springer Science & Business Media, 2013. doi: 10.1007/978-1-4471-5104-3_2.

[97] T. Kacetl, J. Kacetl, N. Tashakor, and S. M. Goetz, "A Simplified Model for the Battery Ageing Potential Under Highly Rippled Load," *IEEE Eur. Conf. Power Electron. (ECCE, EPE)*, vol. 24, pp. 1–10, 2022.